\RequirePackage{ifpdf}
\ifpdf 
\documentclass[pdftex]{sigma}
\else
\documentclass{sigma}
\fi

\begin{document}

\renewcommand{\PaperNumber}{086}

\FirstPageHeading

\renewcommand{\thefootnote}{$\star$}

\ShortArticleName{On Transformations of the Rabelo Equations}

\ArticleName{On Transformations of the Rabelo Equations\footnote{This paper is a
contribution to the Proceedings of the Seventh International Conference ``Symmetry in Nonlinear Mathematical Physics''
(June 24--30, 2007, Kyiv, Ukraine). The full collection is available at
\href{http://www.emis.de/journals/SIGMA/symmetry2007.html}{http://www.emis.de/journals/SIGMA/symmetry2007.html}}}

\Author{Anton SAKOVICH~$^\dag$ and Sergei SAKOVICH~$^\ddag$}

\AuthorNameForHeading{A.~Sakovich and S.~Sakovich}

\Address{$^\dag$~National Centre of Particle and High Energy Physics, 220040 Minsk, Belarus}
\EmailD{\href{mailto:ant.s@tut.by}{ant.s@tut.by}}
\URLaddressD{\url{http://ant.s.at.tut.by}}

\Address{$^\ddag$~Institute of Physics, National Academy of Sciences, 220072 Minsk, Belarus}
\EmailD{\href{mailto:saks@tut.by}{saks@tut.by}}

\ArticleDates{Received May 28, 2007, in f\/inal form August 22, 2007; Published online September 03, 2007}

\Abstract{We study four distinct second-order nonlinear equations of Rabelo which describe pseudospherical surfaces. By transforming these equations to the constant-characteristic form we relate them to some well-studied integrable equations. Two of the Rabelo equations are found to be related to the sine-Gordon equation. The other two are transformed into a linear equation and the Liouville equation, and in this way their general solutions are obtained.}

\Keywords{nonlinear PDEs; transformations; integrability}

\Classification{35Q58; 35Q53; 35C05}

\section{Introduction}

In \cite{R} Rabelo proved that the partial dif\/ferential equation
\begin{gather}
  u_{xt} = \bigl( \bigl( \alpha g(u) + \beta \bigr) u_x \bigr)_{x} \pm g'(u),
	\label{eq1}
\end{gather}
where $g(u)$ is any solution of the linear ordinary dif\/ferential equation
\begin{gather}
	g''(u) + \mu g(u) = \theta ,
	\label{eq2}
\end{gather}
and $\alpha$, $\beta$, $\mu$ and $\theta$ are arbitrary constants, describes pseudospherical surfaces and possesses a~zero-curvature representation with a parameter. Later this equation was studied in \cite{BRT}, where its Lax pair was given explicitly and a B\"{a}cklund self-transformation was constructed.

When $\alpha = 0$, one can easily turn the system of equations \eqref{eq1} and \eqref{eq2} into the well-studied linear equation, Liouville equation or sine-Gordon equation, using appropriate (complex-valued, in general) Galilean and scale transformations of $x$ and $t$ and af\/f\/ine transformations of $u$. The case of a nonzero $\alpha$ is more rich and interesting. When $\alpha \neq 0$, the system of equations~\eqref{eq1} and~\eqref{eq2} is equivalent, through transformations of the above-mentioned type, to one of the following four distinct nonlinear equations (see the Appendix on page \pageref{A} for some details):
\begin{gather}
		u_{xt} = 1 + \tfrac{1}{2}\left(u^2\right)_{xx}, \label{qr}\\
		u_{xt} = u + \tfrac{1}{6}\left(u^3\right)_{xx}, \label{cr}\\
		u_{xt} = \exp u - (\exp u)_{xx}, \label{er}\\
		u_{xt} = \sin u - (\sin u)_{xx}. \label{sr}
\end{gather}
In this paper we refer to equations \eqref{qr}--\eqref{sr} as the Rabelo equations; moreover, in order to distinguish between them, we call equations \eqref{qr}, \eqref{cr}, \eqref{er} and \eqref{sr}, respectively, the quadratic Rabelo equation, cubic Rabelo equation, exp-Rabelo equation and sine-Rabelo equation.

One can derive Lax pairs of equations \eqref{qr}--\eqref{sr} from the Lax pair of system \eqref{eq1}--\eqref{eq2} given in~\cite{BRT}. The quadratic and cubic Rabelo equations \eqref{qr} and~\eqref{cr} are associated with the well-known Wadati--Konno--Ichikawa spectral problem~\cite{WKI}, whereas the spectral problem of the exp-Rabelo and sine-Rabelo equations \eqref{er} and \eqref{sr} is of a dif\/ferent type which, to our knowledge, has not been studied as yet. The main aim of this paper, however, is to show that it is not necessary to apply any special inverse scattering transform technique to the Rabelo equations \eqref{qr}--\eqref{sr} because these equations can be transformed to some well-studied integrable equations. In Section~\ref{s1} we transform the quadratic Rabelo equation to a linear equation, thus obtaining the general solution of equation \eqref{qr} in an implicit form. In Section~\ref{s2} we transform the cubic Rabelo equation \eqref{cr} to the sine-Gordon equation. In Section~\ref{s3} we f\/ind a transformation relating the exp-Rabelo equation to the Liouville equation and obtain in this way the general solution of equation \eqref{er} in a parametric form. In Section~\ref{s4} we f\/ind a transformation relating the sine-Rabelo equation to the sine-Gordon equation. Section~\ref{s5} contains a summary of our results.

Some words are due on what else is known about equations \eqref{qr}--\eqref{sr}. The quadratic Rabelo equation \eqref{qr} is a special case of the Calogero equation $u_{xt} + u u_{xx} + F ( u_x ) = 0$ solvable by quadrature for any function $F$ \cite{C} (note also a dif\/ferent approach \cite{P1} to solve the Calogero equation). The cubic Rabelo equation \eqref{cr} appeared recently in nonlinear optics as the Sch\"{a}fer--Wayne short pulse equation describing the propagation of ultra-short light pulses in silica optical f\/ibers \cite{SW}. The transformation between the cubic Rabelo equation (a.k.a.\ the short pulse equation) and the sine-Gordon equation was discovered in \cite{SS}, and later it was used in \cite{SS2} for deriving exact loop and pulse solutions of equation \eqref{cr} from known kink and breather solutions of the sine-Gordon equation. The recursion operator \cite{SS}, Hamiltonian structures and conserved quantities \cite{B1,B2}, multisoliton solutions \cite{M} and periodic solutions \cite{P2} of equation \eqref{cr} were found and studied as well. On the contrary, the exp-Rabelo and sine-Rabelo equations \eqref{er} and \eqref{sr} have not been studied in any detail as yet, neither have we known about any application of equations \eqref{qr}, \eqref{er} and \eqref{sr}.

\section{The quadratic Rabelo equation} \label{s1}

Let us bring equation \eqref{qr} into a form with constant characteristic directions (consult, e.g., \cite{O} for a def\/inition of characteristics). We make the change of independent variables
\begin{gather}
	u(x,t) = v(y,s): \qquad y=y(x,t), \qquad s=s(x,t)
	\label{new_vars}
\end{gather}
in equation \eqref{qr}, and require that the coef\/f\/icients at $v_{yy}$ and $v_{ss}$ in the resulting expression are equal to zero, which imposes two conditions
\begin{gather}
	y_x \left(y_t - v y_x\right) = 0, \qquad s_x \left(s_t - v s_x\right) = 0
	\label{coef0}
\end{gather}
on two functions $y(x,t)$ and $s(x,t)$ of the transformation. Taking into account the nondegeneracy condition $y_x s_t - y_t s_x \neq 0$, we choose
\begin{gather}
	y_t - v y_x = 0
	\label{qr1_a}
\end{gather}
and $s_x = 0$ in system \eqref{coef0} without loss of generality; moreover, we set $s=t$ for simplicity. As a~result, equation \eqref{qr} in variables \eqref{new_vars} with $s=t$ takes the form
\begin{gather}
  v_{yt} = 1 / y_x ,
	\label{qr1_b}
\end{gather}
where $y(x,t)$ satisf\/ies condition \eqref{qr1_a}. Inverting $y = y(x,t)$ as $x = x(y,t)$, we get $y_x = 1/x_y$ and $y_t = -x_t/x_y$, which brings the system of equations \eqref{qr1_a}--\eqref{qr1_b} into the constant-characteristic form
\begin{gather}
	 v_{yt} - x_y = 0, \qquad x_t + v = 0.
	\label{qr2}
\end{gather}

This linear system \eqref{qr2}, equivalent to the quadratic Rabelo equation \eqref{qr}, can be solved easily. We eliminate $v(y,t)$ from system \eqref{qr2}, get the equation
\begin{gather}
	x_{ytt} + x_y = 0,
\end{gather}
f\/ind its general solution, and in this way obtain the following expressions:
\begin{gather}
  x = a(y) \sin t + b(y) \cos t + c(t), \qquad v = - a(y) \cos t + b(y) \sin t - c'(t),
  \label{qr_s}
\end{gather}
where $a(y)$, $b(y)$ and $c(t)$ are arbitrary functions. These expressions \eqref{qr_s}, where we should replace $v$ by $u(x,t)$, give us a parametric form of the general solution of the quadratic Rabelo equation~\eqref{qr}, the variable $y$ being the parameter. Note, however, that the general solution of equation~\eqref{qr} must contain two arbitrary functions of one variable, whereas we have three arbitrary functions in expressions \eqref{qr_s}. This redundant arbitrariness corresponds to the invariance of system~\eqref{qr2} with respect to an arbitrary transformation $y \to \phi (y)$, which just means that $y$ is a parameter. Note also that, due to $x_y \neq 0$, the functions $a(y)$ and $b(y)$ cannot be simultaneously constant in expressions \eqref{qr_s}, therefore we can always make $a=y$ or $b=y$ by an appropriate transformation $y \to \phi (y)$ of the parameter.

We can rewrite the parametric general solution \eqref{qr_s} of equation \eqref{qr} in the following implicit form. When $a(y) \neq \mathrm{const}$, we make $a=y$, eliminate $y$ from expressions \eqref{qr_s} for $x$ and $u(x,t) = v$, and obtain
\begin{gather}
	\bigl( x - c(t) \bigr) \cos t + \bigl( u + c'(t) \bigr) \sin t = b \bigl( \bigl( x - c(t) \bigr) \sin t - \bigl( u + c'(t) \bigr) \cos t \bigr),
	\label{qr_s1}
\end{gather}
where $b$ and $c$ are arbitrary functions of their arguments. This expression \eqref{qr_s1} is the implicit ``almost general'' solution of equation \eqref{qr}, where the word ``almost'' reminds of the restriction $a(y) \neq \mathrm{const}$ used. The case of $a(y) = \mathrm{const}$ in  expressions \eqref{qr_s} leads to the special solution $u = x \tan t + d(t)$ of equation \eqref{qr}, where $d(t)$ is an arbitrary function; this special solution is not covered by expression \eqref{qr_s1}, and jointly they constitute the general solution of the quadratic Rabelo equation.

We believe that our general solution of equation \eqref{qr} is much simpler than the one derivable by the method used in \cite{C}, though it is very likely that expressions \eqref{qr_s} could be obtained by the method proposed in \cite{P1}.

\section{The cubic Rabelo equation} \label{s2}

Following the same way as in Section \ref{s1}, we make the change of independent variables \eqref{new_vars} in equation \eqref{cr}, require that the coef\/f\/icients at $v_{yy}$ and $v_{ss}$ in the resulting expression are equal to zero, choose $s = t$, invert $y = y(x,t)$ as $x = x(y,t)$, and thus transform the cubic Rabelo equation to the system
\begin{gather}
	v_{yt} - v x_y =0, \qquad x_t + \tfrac{1}{2} v^2 = 0
	\label{cr1}
\end{gather}
possessing constant characteristics. Next we eliminate $x(y,t)$ from equations \eqref{cr1} and integrate the resulting third-order equation
\begin{gather}
  \left( \frac{v_{yt}}{v} \right)_t + v v_y = 0,
\end{gather}
using the integrating multiplier $2v_{yt}/v$. As a result, the cubic Rabelo equation is transformed to the equation
\begin{gather}
  \frac{v_{yt}^{2}}{v^2} + v_y^2 = f(y),
  \label{cr2}
\end{gather}
where $f(y)$ is an arbitrary function.

When $f(y) \neq 0$ in equation \eqref{cr2}, we make $f(y) = 1$, using an appropriate transformation $y \to \phi(y)$, and obtain the equation
\begin{gather}
  v_{yt} = v \big(1 - v_y^2\big)^{1/2},
  \label{cr3}
\end{gather}
where we omit the $\pm$ sign in the right-hand side because the sign can be changed by $y \to -y$. The potential transformation $v = z_t (y,t)$ relates equation \eqref{cr3} to the sine-Gordon equation
\begin{gather}
	z_{yt} = \sin z.
	\label{sg}
\end{gather}
As a result of this, taking into account equations \eqref{cr1}, we obtain the following transformation between the cubic Rabelo equation \eqref{cr} and the sine-Gordon equation \eqref{sg}:
\begin{gather}
  u(x,t) = z_t (y,t), \qquad x = x(y,t): \qquad
  \left\{
  \begin{split}
    x_y &= \cos z ,\\
    x_t &= - \tfrac{1}{2} z_t^2 .
  \end{split}
  \right.
  \label{cr_tr}
\end{gather}
This transformation was found for the f\/irst time in \cite{SS}, in a dif\/ferent way and in a more involved form; later, in \cite{SS2}, it was brought into the present form and used for deriving exact solitary wave solutions of equation \eqref{cr}. Note that, for any solution $z(y,t)$ of the sine-Gordon equation, transformation \eqref{cr_tr} produces a solution $u(x,t)$ of the cubic Rabelo equation, determined up to an arbitrary constant shift of $x$ and given in a parametric form, with $y$ being the parameter.

When $f(y) = 0$ in equation \eqref{cr2}, we get $v_y = \pm i x_y$ due to the f\/irst equation of system \eqref{cr1}, which immediately leads to the special solution
\begin{gather}
  u = \pm i x + g(t)
  \label{cr_s}
\end{gather}
of the cubic Rabelo equation, where $g(t)$ is an arbitrary function. This expression \eqref{cr_s} embraces all those solutions of equation \eqref{cr} which cannot be obtained by means of transformation \eqref{cr_tr}.

\section{The exp-Rabelo equation} \label{s3}

Using again the method described in Section \ref{s1}, we transform the exp-Rabelo equation \eqref{er} to the following constant-characteristic system of equations:
\begin{gather}
	v_{yt} - x_y \exp v = 0, \qquad x_t - \exp v = 0.
  \label{er1}
\end{gather}
Next we eliminate $v(y,t)$ from equations \eqref{er1}, make for simplicity the substitution $x(y,t) = \log w(y,t)$, integrate the resulting equation
\begin{gather}
  \left( \frac{w_{yt}}{w_t} \right)_t - \frac{w_{yt}}{w} = 0 ,
\end{gather}
using the integrating multiplier $w_t/w_{yt}$, and in this way obtain
\begin{gather}
  w_{yt} = f(y) w w_t,
  \label{er2}
\end{gather}
where $f(y)$ is an arbitrary function.

When $f(y) \neq 0$ in equation \eqref{er2}, we make $f(y) = 1$ by an appropriate transformation $y \to \phi (y)$. In this case the potential transformation $w = z_y (y,t)$ relates equation \eqref{er2} to the Liouville equation
\begin{gather}
  z_{yt} = \exp z.
  \label{l}
\end{gather}
Simplifying the chain of transformations used, we f\/ind that the exp-Rabelo equation \eqref{er} is related to the Liouville equation \eqref{l} as follows:
\begin{gather}
	x = \log z_y , \qquad u(x,t) = z - \log z_y .
	\label{er_tr}
\end{gather}
Then, using transformation \eqref{er_tr} and the well-known general solution
\begin{gather}
	z = \log \frac{2a'(y) b'(t)}{\bigl( a(y) + b(t) \bigr)^2}
\end{gather}
of the Liouville equation \eqref{l}, where $a(y)$ and $b(t)$ are arbitrary functions, we obtain the parametric ``almost general'' solution
\begin{gather}
	x = \log \left( \frac{a''(y)}{a'(y)} - \frac{2a'(y)}{a(y) + b(t)} \right), \qquad
	u(x,t) = - x + \log \frac{2a'(y) b'(t)}{\bigl( a(y) + b(t) \bigr)^2}
  \label{er_s}
\end{gather}
of the exp-Rabelo equation \eqref{er}, with $y$ being the parameter, where the word ``almost'' reminds of the restriction $f(y) \neq 0$ used. Evidently, it is impossible to rewrite this parametric solution in an implicit form by eliminating $y$ from expressions \eqref{er_s} with arbitrary functions $a(y)$ and $b(t)$.

When $f(y)=0$, equation \eqref{er2} is linear, and we obtain the special solution
\begin{gather}
	u = -x + g(t)
\end{gather}
of equation \eqref{er}, where $g(t)$ is an arbitrary function; this special solution is not covered by the ``almost general'' solution \eqref{er_s}, and jointly they constitute the general solution of the exp-Rabelo equation.

\section{The sine-Rabelo equation} \label{s4}

Doing all the same as in Section \ref{s1}, we bring the sine-Rabelo equation \eqref{sr} into the constant-characteristic form
\begin{gather}
	v_{yt} - x_y \sin v = 0, \qquad x_t - \cos v = 0.
  \label{sr1}
\end{gather}
Next we eliminate $x(y,t)$ from system \eqref{sr1}, integrate the resulting equation
\begin{gather}
	\left( \frac{v_{yt}}{\sin v} \right)_t - (\cos v)_y = 0 ,
\end{gather}
using the integrating multiplier $2v_{yt} / \sin v$, and in this way obtain
\begin{gather}
	\left( \frac{v_{yt}}{\sin v} \right)^2 + v_y^2 = f(y),
	\label{sr2}
\end{gather}
where $f(y)$ is an arbitrary function.

When $f(y) \neq 0$ in equation \eqref{sr2}, we make $f(y) = 1$ by an appropriate transformation $y \to \phi (y)$ and obtain the equation
\begin{gather}
	v_{yt} = \left( 1 - v_y^2 \right)^{1/2} \sin v ,
	\label{sr3}
\end{gather}
where the $\pm$ sign in the right-hand side is omitted because it can be inverted by $y \to -y$. This equation \eqref{sr3} is the modif\/ied sine-Gordon equation of Chen \cite{HC} related to the sine-Gordon equation \eqref{sg} by the transformation $z = v + \arcsin v_y$. Note that we can express the transformation between equations \eqref{sg} and \eqref{sr3} as the B\"{a}cklund transformation
\begin{gather}
	v_y + \sin (v - z) = 0, \qquad v_t - z_t + \sin v = 0.
	\label{c_sg}
\end{gather}
Furthermore, from equations \eqref{sr1} and \eqref{sr3} we have
\begin{gather}
	x_y = \left( 1 - v_y^2 \right)^{1/2} , \qquad x_t = \cos v .
	\label{x_tr}
\end{gather}
These relations \eqref{c_sg} and \eqref{x_tr} completely determine the transformation between the sine-Rabelo equation \eqref{sr} and the sine-Gordon equation \eqref{sg}, which works in the following way. For any given solution $z(y,t)$ of the sine-Gordon equation, relations \eqref{c_sg} determine a corresponding function $v(y,t)$ up to one arbitrary constant (note that it is necessary to solve a f\/irst-order ordinary dif\/ferential equation at this step). Then, with this function $v(y,t)$ obtained, relations \eqref{x_tr} determine a corresponding function $x(y,t)$ up to an arbitrary additive constant of integration. Finally, expressions $u(x,t) = v(y,t)$ and $x = x(y,t)$ determine a solution $u(x,t)$ of the sine-Rabelo equation in a parametric form, with $y$ being the parameter.

When $f(y) = 0$ in equation \eqref{sr2}, we have $v_y = \pm i x_y$ owing to the f\/irst equation of system~\eqref{sr1} and obtain the special solution
\begin{gather}
	u = \pm i x + g(t)
	\label{sr_s}
\end{gather}
of the sine-Rabelo equation, where $g(t)$ is an arbitrary function. This expression \eqref{sr_s} embraces all those solutions of equation \eqref{sr} which cannot be obtained using transformation \eqref{c_sg}--\eqref{x_tr}.

\section{Conclusion} \label{s5}

In this paper we studied four second-order nonlinear equations \eqref{qr}--\eqref{sr} introduced by Rabelo. We found the transformations relating the Rabelo equations to some well-studied integrable equations. The quadratic Rabelo equation \eqref{qr} and the exp-Rabelo equation \eqref{er} were transformed into a linear equation and the Liouville equation, respectively, and in this way their general solutions were obtained. The cubic Rabelo equation \eqref{cr} and the sine-Rabelo equation \eqref{sr} were found to be related to the sine-Gordon equation. We believe that these results can be important for further studies on properties and solutions of the Rabelo equations, especially if there appear some applications of equations \eqref{qr}, \eqref{er} and \eqref{sr} to physics and technology.

\pdfbookmark[1]{Appendix: Four Rabelo equations}{appendix}
\section*{Appendix: Four Rabelo equations} \label{A}

It is a commonplace of soliton theory, and really an easy exercise in the case of system \eqref{eq1}--\eqref{eq2}, to bring an integrable equation under investigation into a form containing no free parameters, using appropriate point transformations of variables. The transformation
\begin{gather*}
	u = c_1 U(X,T) + c_2 , \qquad X = c_3 x + c_4 t , \qquad T = c_5 t
\end{gather*}
with appropriately chosen constants $c_1 , \dotsc , c_5$ ($c_1 c_3 c_5 \neq 0$) is suf\/f\/icient to bring any nonlinear equation of class \eqref{eq1}--\eqref{eq2} with $\alpha \neq 0$ into one of the four distinct forms \eqref{qr}--\eqref{sr} for $U(X,T)$.

When $\mu = 0$ in equation \eqref{eq2}, we have
\begin{gather*}
	g(u) = \tfrac{1}{2} \theta u^2 + \gamma u + \delta
\end{gather*}
with arbitrary constants $\gamma$ and $\delta$. If $\theta = 0$ and $\gamma \neq 0$, we choose
\begin{gather*}
	c_1 = \epsilon / \sqrt{\epsilon \alpha} , \qquad c_2 = 0 , \qquad c_3 = 1 , \qquad c_4 = \alpha \delta + \beta , \qquad c_5 = \gamma \sqrt{\epsilon \alpha} ,
\end{gather*}
where $\epsilon = \pm 1$ corresponds to the $\pm$ sign in equation \eqref{eq1}, and obtain equation \eqref{qr} for $U(X,T)$. If $\theta \neq 0$, we choose
\begin{gather*}
	c_1 = 1 / \sqrt{\epsilon \alpha} , \qquad c_2 = - \gamma / \theta , \qquad c_3 = 1 , \qquad c_4 = \alpha \delta - \alpha \gamma^2 / ( 2 \theta ) + \beta , \qquad c_5 = \epsilon \theta
\end{gather*}
and obtain equation \eqref{cr} for $U(X,T)$. The linear case $\theta = \gamma = 0$ is covered by the case $\alpha = 0$.

When $\mu \neq 0$ in equation \eqref{eq2}, we have
\begin{gather*}
	g(u) = \gamma \exp ( \xi u ) + \delta \exp ( - \xi u ) + \theta / \mu
\end{gather*}
with arbitrary constants $\gamma$ and $\delta$, where $\xi^2 = - \mu$. If $\delta = 0$ and $\gamma \neq 0$, we choose
\begin{gather*}
	c_1 = 1 / \xi , \qquad c_2 = 0 , \qquad c_3 = \xi / \sqrt{- \epsilon \alpha} , \\
	c_4 = ( \beta \xi - \alpha \theta / \xi ) / \sqrt{- \epsilon \alpha} , \qquad c_5 = \epsilon \gamma \xi \sqrt{- \epsilon \alpha}
\end{gather*}
and obtain equation \eqref{er} for $U(X,T)$. The case of $\gamma = 0$ and $\delta \neq 0$ reduces to the above one with $\xi \mapsto - \xi$ and $\gamma \mapsto \delta$. The linear case $\gamma = \delta = 0$ is covered by the case $\alpha = 0$. Finally, if $\gamma \delta \neq 0$, we choose
\begin{gather*}
	c_1 = i / \xi , \qquad c_2 = \log ( \delta / \gamma ) / ( 2 \xi ) , \qquad c_3 = \xi / \sqrt{- \epsilon \alpha} , \\
	c_4 = ( \beta \xi - \alpha \theta / \xi ) / \sqrt{- \epsilon \alpha} , \qquad c_5 = 2 \epsilon \gamma \xi \sqrt{\delta / \gamma} \sqrt{- \epsilon \alpha}
\end{gather*}
and obtain equation \eqref{sr} for $U(X,T)$.

\subsection*{Acknowledgements}

The authors are grateful to Professors N.H.~Ibragimov, A.V.~Mikhailov and P.J.~Olver for their comments on equation \eqref{sr3}, and to the anonymous referee who suggested reference \cite{P1}. A.S.~is grateful to the Organizing Committee of the Seventh International Conference ``Symmetry in Nonlinear
Mathematical Physics'' (June 24--30, 2007, Kyiv) and ICTP Of\/f\/ice of External Acti\-vi\-ties for travel and local support in participating the conference. The work of S.S. was supported in part by the Belarusian Republican Foundation for Fundamental Research.

\pdfbookmark[1]{References}{ref}
\LastPageEnding

\end{document}